\documentstyle[11pt,fleqn]{article}

\begin{document}

\title{Pad\'e Approximants and the Fixed-Points of the $n_f = 3$
QCD $\beta$-Function
}
\smallskip
\author{
V. Elias\\
Department of Applied Mathematics\\
The University of Western Ontario\\
London, Ontario  N6A 5B7\\
Canada\\
and
T. G. Steele\\
Department of Physics and Engineering Physics\\
University of Saskatchewan\\
Saskatoon, Saskatchewan  S7N 5C6\\
Canada
}

\maketitle

\begin{center}
PACS:  12.38.Aw, 11.10.Hi
\end{center}

\smallskip

\begin{abstract}
The positive zeros of $[2|1]$, $[1|2]$ and the most general possible
$[2|2]$ Pad\'e approximants whose Maclaurin series reproduce the
presently known terms in the three-flavour ($n_f = 3$) QCD $\beta$-
function are all shown
to correspond to {\it ultraviolet} fixed points.
\end{abstract}

\newpage

Higher order terms of the QCD $\beta$-function
\renewcommand{\theequation}{1\alph{equation}}
\begin{equation}
\mu^2 \frac{dx}{d \mu^2} \equiv \beta (x), 
\end{equation}
\begin{equation}
\beta(x) = - \sum_{i=0}^\infty \beta_i x^{i+2},
\end{equation}
\renewcommand{\theequation}{\arabic{equation}}
\setcounter{equation}{1}
\noindent $[x \equiv \alpha_s (\mu) / \pi]$ are known to permit the 
occurrence
of fixed points other than the ultraviolet fixed point at $x = 0$;
{\it e.g.} the positive infrared fixed point (IRFP) which occurs for $9 
\leq
n_f \leq 16$ when the series for $\beta(x)$ in (1) is truncated after
two terms $[\beta_0 = (11 - 2n_f / 3) / 4; \;  \beta_1 = (102 - 38n_f
/ 3) / 16; \;  x_{_{IRFP}} = -\beta_0 / \beta_1]$.  However, the fixed
points arising from such truncation are likely to be spurious,
as the candidate-value for $x_{_{IRFP}}$ is sufficiently large for the
highest-order term in the series $\beta(x_{_{IRFP}})$ to be comparable in
magnitude to lower order terms [{\it e.g.} $| \beta_1 x^3 | = | \beta_0 x^2
|$]. In a recent paper [1], Ellis, Karliner and Samuel predicted the
coefficient $\beta_3$ via Pad\'e approximant methods, and argued that
$\beta_{0-2}$ and their prediction for $\beta_3$ yield a Pad\'e
summation of the $\beta$-function [by which the infinite series in
(1b) is not truncated,but matched to an appropriately chosen Pad\'e
approximant] with a nonzero IRFP that is consistent with an earlier
prediction by Mattingly and Stevenson [2]. Such Pad\'e summations
provide a more accurate representation of the full beta function
by estimating the  summation of
the (unknown) higher-order terms [3].

In the present letter, we study the zeros of Pad\'e-approximant
summations of the $n_f = 3$ $\beta$-function that are consistent with
the (now fully-known [4]) coefficients
$\beta_{0-3}$.  In particular, we consider the explicit $[2|1]$ and $[1|2]$
Pad\'e approximants, as well as the most general possible $[2|2]$
approximant, whose Maclaurin expansions reproduce $\beta_{0-3}$.  In
each case, we find that the positive zeros
of the Pad\'e approximant correspond to ultraviolet fixed points and
{\it not} to infrared fixed points.

The known coefficients of the $n_f = 3$ $\beta$-function can be
expressed as follows [4]:
\begin{equation}
\beta(x) = -(9x^2 / 4) [1 + R_1 x + R_2 x^2 + R_3 x^3 + R_4 x^4 +
.],
\end{equation}
with $R_1 = 16/9, \; R_2 = 4.471065, \; R_3 = 20.99027$. 
$R_4$ and subsequent terms are not presently known.  The $[2|1]$
Pad\'e approximant that successfully matches the first four terms in
the series (2) is $(1 - 2.91691x - 3.87504x^2) / (1 - 4.69468x)$. 
The only positive numerator root is at $x = 0.2559$.  This fixed
point, however, is separated from the $\beta$-function's ultraviolet
fixed point (UVFP) at $x = 0$ by a smaller positive zero of the
denominator ($x = 0.2130$).  Consequently, if one uses the $[2|1]$
approximant to represent the $\beta$-function (2), one finds that the
$\beta$-function has negative slope as one approaches $x = 0.2559$
from above, and positive slope as one approaches $x = 0.2559$ from
below.  This behaviour characterizes $0.2559$ as an UVFP.

Such behaviour --- specifically, a positive zero of the denominator
that is less than the first positive zero of the numerator ---
characterizes the $[1|2]$ Pad\'e approximant as well:  $(1 -
8.17337x)/(1 - 9.95115x + 13.2199x^2)$.  The numerator zero at $x =
0.1223$ is separated from the $x = 0$ UVFP by a denominator zero at $x =
0.1194$.  Thus the zero of the $[1|2]$ Pad\'e approximant 
that matches the known terms of the series in (2) again corresponds to
a UVFP of the $\beta$-function (2).

The $R_4$ term of the series (2) can be estimated using an algorithm
[5] based upon the asymptotic error formula [1,6] relating
the value $R_{N+2}$ to the predicted value $R_{N+2}^{\rm Pad\acute{e}}$
obtained from expanding an $[N|1]$ Pad\'e approximant into a
Maclaurin series:
\begin{equation}
\delta_{N+2} \equiv \frac{R_{N+2}^{\rm Pad\acute{e}} - R_{N+2}}{R_{N+2}} =
\frac{-A}{[N+1+(a+b)]}.
\end{equation}
Using a $[0|1]$ approximant, one finds that $\delta_2 =
(R_1^2 - R_2) / R_2  =  -A / [1+(a+b)]$.  Using a $[1|1]$
approximant, one finds that $\delta_3 = (R_3^2/R_2 - R_3)/R_3 = -A /
[2 + (a+b)]$.  Since $R_{1,2,3}$ are known, these two relations
determine the two unknowns $A$ and $(a+b)$.  One can estimate the unknown
coefficient $R_4$ by applying (3) to the $[2|1]$ approximant:
\begin{equation}
R_4 = \frac{R_3^2 / R_2}{1 + \delta_4} = \frac{R_3^2(R_2^3 + R_1 R_2
R_3 - 2R_1^3 R_3)}{R_2(2R_2^3 - R_1^3 R_3 - R_1^2 R_2^2)}.
\end{equation}
For the $n_f = 3$ values of $R_{1,2,3}$ given above, we
find $R_4 = -849.7$.

Using these numbers, the polynomial $1+R_1 x + R_2 x^2 + R_3 x^3 +
R_4 x^4$ {\it does} have a positive zero which can be identified with a
$\beta$-function IRFP at $x = 0.2143 \; (\alpha_s = 0.673)$,
provided we accept a degree-4 truncation of the series in (2).  Such
an IRFP [analogous to the naive IRFP $x = -\beta_1 / \beta_0$
described at the beginning of this letter] is of questionable
validity because of the large magnitude of the dominant $R_4 x^4$ term
immediately preceding truncation [7].

Such truncation difficulties are averted if the known coefficients
$R_{1,2,3}$ and the estimated coefficient $R_4$ are utilized to
generate a $[2|1]$ Pad\'e approximant that reproduces $1 + R_1 x +
R_2 x^2 + R_3 x^3 + R_4 x^4$ as the first five terms of its infinite
Maclaurin series.  This approximant, $(1 + 94.383x - 75.605x^2) / (1
+ 92.606x - 244.71 x^2)$, has one positive numerator-zero $(x =
1.259)$, which is found to be larger than the only positive
denominator-zero $(x = 0.3889)$.  Consequently, the positive zero of
the $[2|2]$ approximant generated via the estimate (4) for $R_4$ once
again corresponds to a UVFP of the $\beta$-function (2).

Surprisingly, this correspondence holds even if we discard (4)
entirely and
develop a general $[2|2]$ Pad\'e approximant whose $R_4$ dependence
is explicit [8].  Using $R_{1-3}$ appropriate for the $n_f = 3$
$\beta$-function, one finds the most general $[2|2]$ approximant 
whose Maclaurin
series reproduces $1 + R_1 x + R_2 x^2 + R_3 x^3 + R_4 x^4$ with
$R_4$ arbitrary to be $(1 + a_1 x + a_2 x^2) / (1 + b_1 x + b_2 x^2)$,
such that $a_1 = 7.1945 - 0.10261 R_4$, $a_2 = -11.329 + 0.075643
R_4$, $b_1 = 5.4168 - 0.10261 R_4$, $b_2 = -25.430 + 0.25806
R_4$.  The numerator and denominator zeros are (respectively) denoted
by $x_{\pm} \equiv (-a_1 \pm \sqrt{a_1^2 - 4a_2}) / 2a_2$,
$y_{\pm} \equiv (-b_1 \pm \sqrt{b_1^2 - 4b_2}) / 2b_2$.  For $R_4 <
98.54$, both $a_2$ and $b_2$ are negative, in which case $x_-$ and
$y_-$ are positive, and $x_+$ and $y_+$ are negative.  Fig. 1 shows
that $0 < y_- < x_-$ through this range, in which case the positive
root $x_-$ necessarily corresponds to a UVFP.  For $R_4$ between
$98.54$ and $149.76$, $x_-$, $y_+$ and $y_-$ are all positive ($x_+$
is negative).  Noting that $y_+ > y_-$ in this range,one finds that
the numerator zero can be an IRFP if either $y_+ > y_- > x_- > 0$, 
or $x_- > y_+ > y_- > 0$.  Neither of these sets of inequalities is 
upheld over this
range of $R_4$.  Instead $y_+ > x_- > y_- > 0$ [Fig. 1], consistent
with $x_-$ corresponding to a UVFP of the $\beta$-function.  Finally,
if $R_4 > 149.76$, we see that $x_+ > x_- > 0$ and $y_+ > y_- > 0$. 
However, these four positive roots are seen to satisfy $x_+ > y_+ >
x_- > y_- > 0$ [Fig. 1], consistent with identifying {\it both} zeros
of the Pad\'e-approximant numerator with UVFP's of the $n_f = 3$
$\beta$-function.  Corresponding behaviour of the coupling constant
is heuristically presented in Fig 2.  This same behaviour has already been shown
to characterize the {\em exact} $\beta$ function for SUSY gluodynamics [10].

Thus, no matter what $\beta_4$ ($=\beta_0 R_4$) is eventually found
to be, the $[2|2]$ Pad\'e approximant whose Maclaurin expansion
matches the $\beta_{0-4}$ terms of the $\beta$-function will not
support the existence of any positive IRFPs; zeros of this
approximant all correspond to UVFPs.  As is evident from Fig. 2, the
structure of the $[2|2]$ approximant to the $\beta$-function (2)
decouples the IR-region entirely from coupling-constant evolution
between UVFPs --- {\it i.e.} if $x$ is between zero and $x_-$, $\mu$ cannot
be smaller than $\mu(y_-)$.  Finally we note that the existence of a
UVFP different from zero [necessarily leading to a double-valued
function for $\alpha_s (\mu)$] could indicate an additional 
strong-coupling
phase of QCD at short distances [9], with possible implications for
dynamical electroweak symmetry breaking.  QCD may conceivably furnish 
its own
technicolour.

\section*{Acknowledgements}

We are grateful to the Natural Sciences and Engineering Research
Council of Canada for support, and to A. S. Deakin, R. Migneron and
V. A. Miransky for related discussions.  We also wish to acknowledge
the late Mark Samuel for taking the time to acquaint us with Pad\'e
approximant methods and to familiarize us with his extensive work in
this area;  his sudden passage is a loss deeply felt.

\section*{Figure Captions:}
\begin{description}
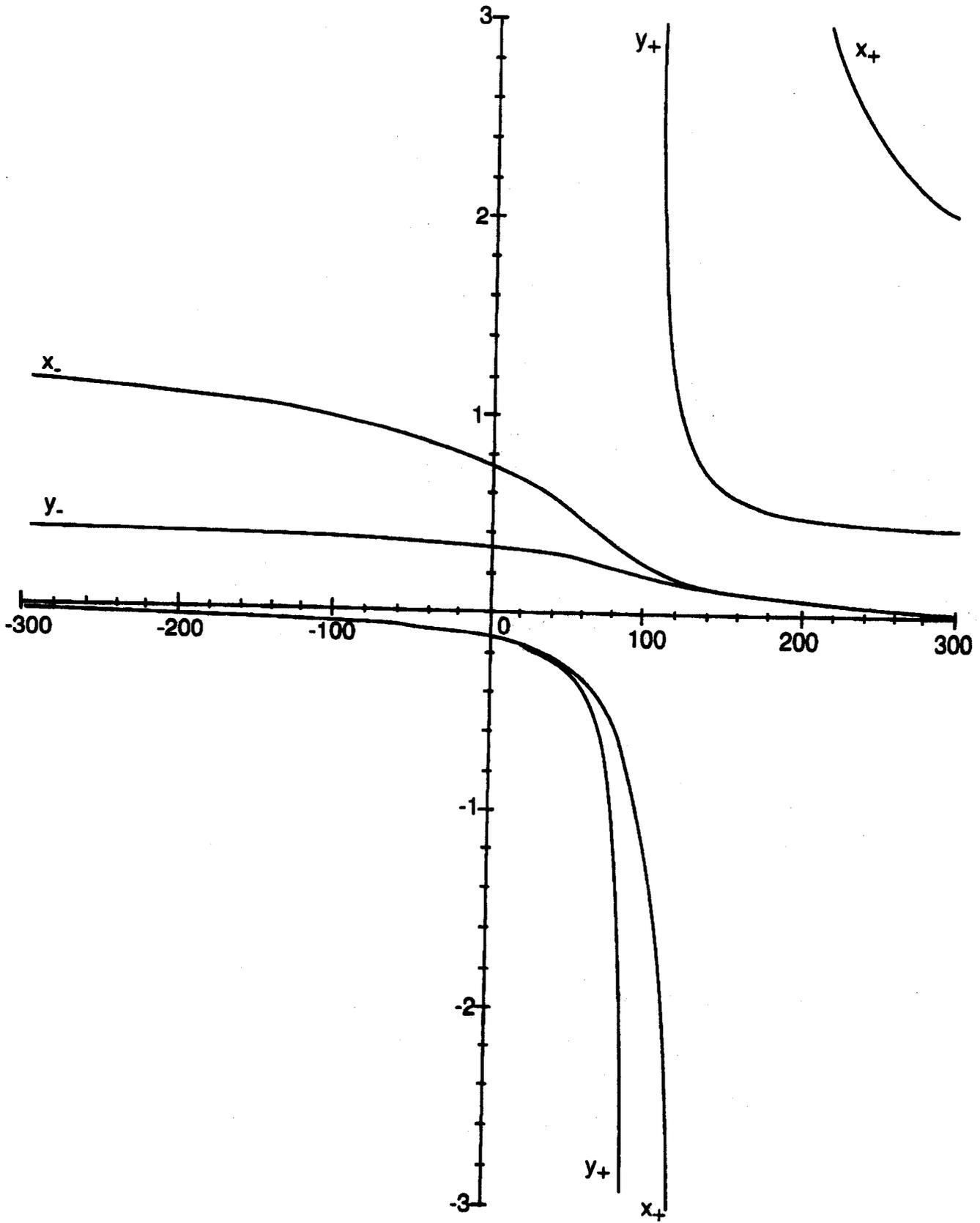
\item{Figure 1:}  Relative size of eq.(4)'s numerator zeros $x_\pm$ and 
denominator 
zeros $y_\pm$, expressed as functions of the horizontal-axis variable 
$R_4$.
$x_-$ approaches $y_-$ from above for large positive values of $R_4$.
\end{description}
\begin{description}
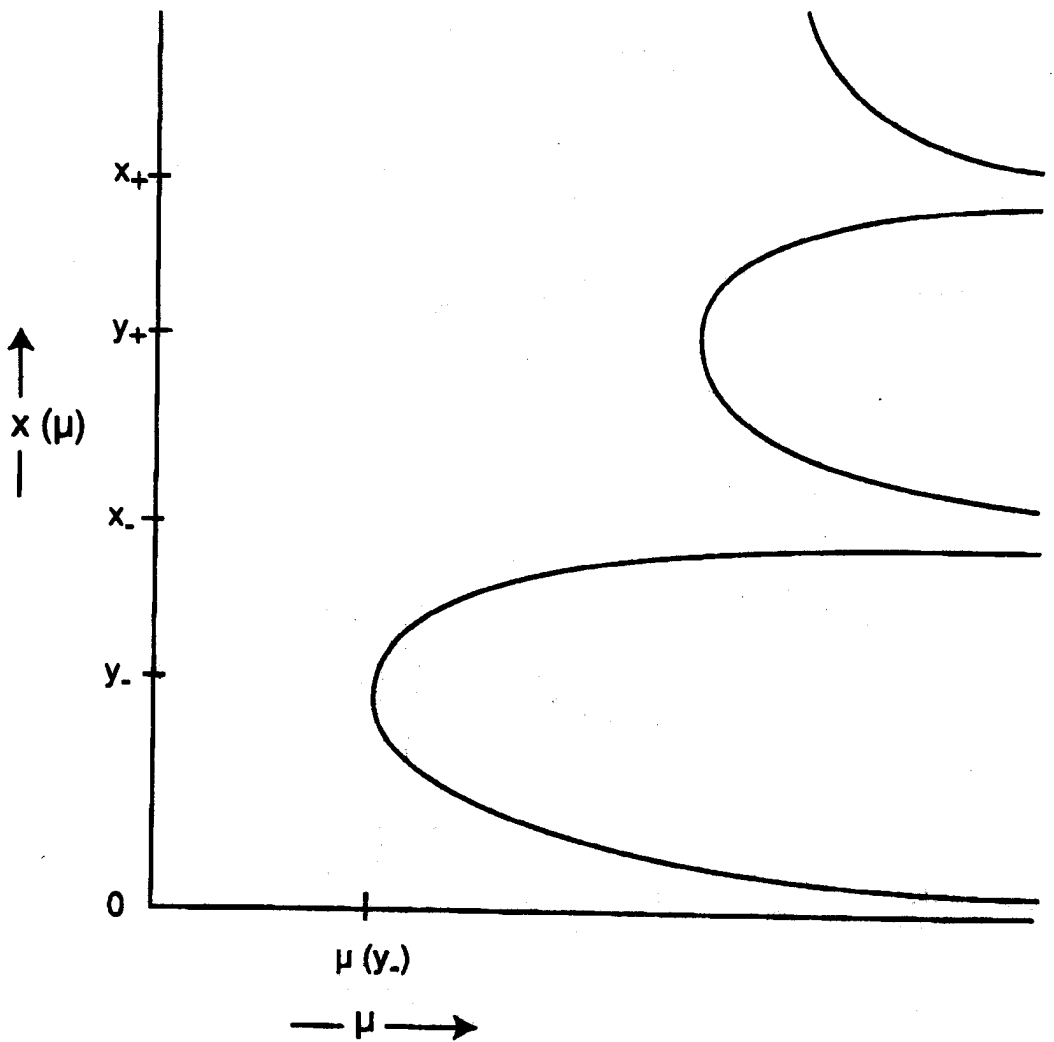
\item{Figure 2:}  Schematic behaviour of $x(\mu)$ obtained from use of the
$[2|2]$ Pad\'e approximant for $x_+ > y_+ > x_- > y_- > 0$. $x_+$ and $x_-
$ 
are numerator zeros corresponding to UVFP's. Corresponding behaviour
of (positive) $x(\mu)$ when $x_- > y_- > 0$ with $x_+, y_+$ both
negative (see Fig. 1) is obtained by excising the middle branch of
the above figure.  The value of $\mu$ when $x = y_-$, the first zero of 
the Pad\'e-denominator, is denoted by $\mu(y_-)$.
\end{description}

\end{document}